\documentclass[epj]{svjour}
\usepackage{graphics}
\begin{document}
\title{Unified description of the coupled-channels and statistical Hauser-Feshbach nuclear reaction theories for low energy neutron incident reactions}
\author{Toshihiko Kawano\inst{1}}
\institute{Theoretical Division, Los Alamos National Laboratory, Los Alamos, NM 87545, USA, \email{kawano@lanl.gov}}
\date{Received: date / Revised version: date}
\abstract{ We incorporate the coupled-channels optical model into the
  statistical Hauser-Feshbach nuclear reaction theory, where the
  scattering matrix is diagonalized by performing the
  Engelbrecht-Weidenm\"{u}ller transformation. This technique
  has been implemented in the coupled-channels optical model code ECIS
  by J.~Raynal, and we extend this method so that all the open channels in
  a nucleon-induced reaction on a deformed nucleus can be calculated
  consistently.
} 

\maketitle

\section{Introduction}
\label{sec:introduction}
It is well known that the nucleon-induced scattering process from a
nucleus can be described by the optical model, where the imaginary
part of the potential represents a deficit of incoming flux, hence the
scattering $S$-matrix of the optical model is no longer unitary.  At
relatively low incident energies, where only a few open channels are
involved, an absorbed particle once forms a compound state in the
target nucleus, and it comes back to the entrance channel as the
compound elastic scattering process. The total wave-function is the
coherent sum of the incoming and out-going
waves~\cite{Moldauer1961,Moldauer1964}. Although the direct (or
``shape'') elastic scattering $(d\sigma/d\Omega)^{\rm SE}$ and the
compound elastic scattering $(d\sigma/d\Omega)^{\rm CE}$ cannot be
distinguished experimentally, theoretical interpretation divides these
two scattering processes into different time-scale domains --- the
fast and slow parts. The optical model gives the faster part, while
the statistical model for nuclear
reactions~\cite{Kawai1973,Hofmann1975,Moldauer1980,Verbaarschot1985,Kawano2015}
accounts for the slower compound elastic scattering process.  We
understand the experimental elastic scattering data are the incoherent
sum of both processes. The optical model codes, which have been
utilized to analyze the nucleon scattering data, for example
ECIS~\cite{Raynal1972}, ELIESE-3~\cite{JAERI-1224},
CASTHY~\cite{JAERI-1321}, ABAREX~\cite{ANLNDM145}, and so forth, are capable for
calculating the compound nucleus (CN) process, yet they are limited to a
binary reaction with relatively small channel space. Note that the
aforementioned codes are just examples, and there exist more computer
programs that calculate the optical model in the nuclear science
field.

As a general descriptions of nuclear reaction process, the
Hauser-Feshbach (HF) statistical theory~\cite{Hauser1952} is extended
to a multi-stage reaction, where a residual nucleus formed after
particle emission is allowed to further decay as another CN
process. In this case, the main part of the calculation is the
compound nuclear reaction, and the optical model is somewhat hidden
behind it. The so-called HF codes, GNASH~\cite{LA6947,LA12343},
TNG~\cite{ORNL-7042}, STAPRE~\cite{IRK-76}, EMPIRE~\cite{INDC0603},
TALYS~\cite{Koning2004,Koning2012} and so on, invoke an optical model
code to generate transmission coefficients $T_{lj}$ as a model input,
where the quantum numbers $lj$ are the orbital angular momentum and
the spin. J.~Raynal's ECIS code has been widely used for this
purpose. Exceptions are the CCONE~\cite{Iwamoto2007,Iwamoto2013} and
CoH$_3$~\cite{Kawano2010,Kawano2019} codes, those include a private
$T_{lj}$ generator internally.

These HF codes, despite they are capable for calculating nuclear
reaction cross sections for all the open channels, do not furnish a
strict connection between the optical and statistical models.  This
issue becomes more serious when the target nucleus is strongly
deformed, and the single channel optical model has to be extended to
the coupled-channels (CC) formalism~\cite{Tamura1965}. There are two
vague approximations made by these codes to deal with the nuclear
deformation effect; (a) $T_{lj}$ for the excited states are replaced
by the one for the ground state by correcting the channel energy, and
(b) the direct reaction effect in the statistical
theory~\cite{Hofmann1975,Engelbrecht1973,Moldauer1975b} is
ignored. Raynal carefully dealt with these issues in ECIS, and
established a unique integration of the optical and HF statistical
models, nevertheless the formalism was limited to the binary
reactions only.

Like ECIS or ELIESE-3, CoH was originally developed as a nucleon
scattering data analysis code~\cite{Kawano1999c}, which includes the
statistical HF theory with the width fluctuation correction by
Moldauer~\cite{Moldauer1980}. Later, CoH was extended to the full
multi-stage HF code, so that the optical model and the statistical
model were naturally unified just like ECIS, but in more general
sense. In this paper we present the unified description of CC
optical and HF statistical models implemented in the CoH$_3$ code,
where Raynal's ideas in ECIS were important resources and clues.  We
demonstrate how the approximations made by the existing HF codes bring
systematic uncertainties in the predicted cross sections. We limit
ourselves to the low energy neutron induced reactions only, where the
compound elastic scattering plays more important role than the
charged-particle cases.

\section{Theoy}
\label{sec:theory}
\def\ave#1{\left\langle{#1}\right\rangle}

\subsection{Compound nuclear reaction and transmission coefficient}

The energy-averaged cross section for a reaction from channel $a$
to channel $b$ is written by the partial decay width $\Gamma_c$ as
\begin{equation}
  \sigma^{\rm CN}_{ab}
  = \frac{\pi}{k_a^2} g_a
    \frac{2\pi}{D}
    \left\langle
       \frac{\Gamma_a \Gamma_b}{\sum_c \Gamma_c}
    \right\rangle \ ,
  \label{eq:CN}
\end{equation}
where $D$ is the average resonance spacing, $k_a$ is the wave number
of incoming particle, and $g_a$ is the spin factor given later.  By
applying a relation between the single-channel transmission
coefficient $T_a$ and the decay width $\Gamma_a$,
\begin{equation}
  T_a \simeq 2\pi\frac{\langle\Gamma_a\rangle}{D} \ ,
\end{equation}
the width fluctuation corrected HF cross
section~\cite{Moldauer1961,Hauser1952} reads
\begin{eqnarray}
  \sigma^{\rm CN}_{ab}
   &=& \frac{\pi}{k_a^2} g_a
       \frac{T_a T_b}{\sum_c T_c} W_{ab} \nonumber \\
   &=& \sigma^{\rm HF}_{ab} W_{ab} \ ,
   \label{eq:HF}
\end{eqnarray}
where $\sigma^{\rm HF}_{ab}$ is the original HF cross section.  The
width fluctuation correction factor $W_{ab}$ is also a function of
$T_a$. When the resonance decay width $\Gamma_a$
forms the $\chi^2$ distribution with the channel degree-of-freedom
$\nu_a$, the width fluctuation correction factor can be evaluated
numerically as~\cite{Moldauer1975b,Moldauer1975a,Moldauer1976}
\begin{eqnarray}
    W_{ab} &=& \left(1+\frac{2\delta_{ab}}{\nu_a}\right)
   \int_0^\infty\!\!\!
   \frac{dt}{F_a(t)F_b(t) \prod_k F_k(t)^{\nu_k/2}} ,
     \label{eq:MoldauerW}  \\
  F_k(t) &=& 1 + \frac{2}{\nu_k} \frac{T_k}{\sum_c T_c} t \ .
\end{eqnarray}
The Gaussian Orthogonal Ensemble (GOE) model~\cite{Verbaarschot1985}
provides accurate estimates of $\nu_a$~\cite{Moldauer1980,Kawano2015}
in terms of $T_a$, and all of the reaction cross sections
$\sigma_{ab}$ are determined by $T_a$ and $T_b$ only.

\begin{figure}
  \begin{center}
    \resizebox{0.7\columnwidth}{!}{\includegraphics{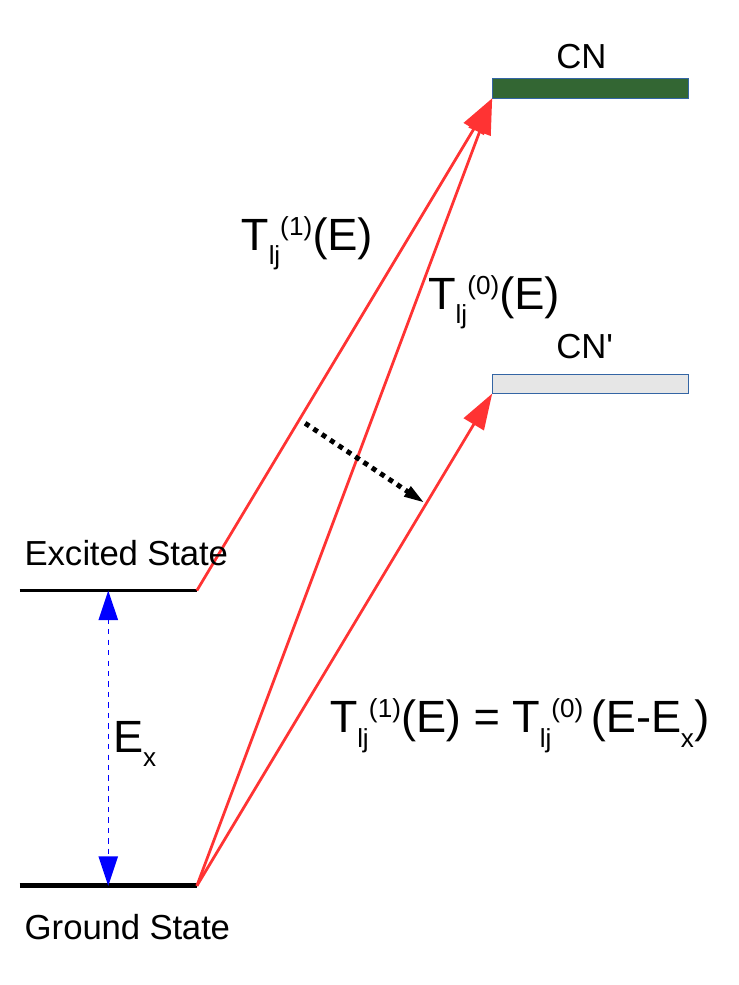}}
   \end{center}
  \caption{The ground and excited states in a target nucleus and the
    compound nucleus (CN) connected by the particle transmission
    coefficients, $T_{lj}^{(0)}$ and $T_{lj}^{(1)}$. $T_{lj}^{(1)}$ is often
    approximated by shifting to the ground state, as shown by the dotted arrow.}
  \label{fig:transmissionSOM}
\end{figure}

When the ground state of nucleus does not couple so strongly with other
states by collective excitation, the scattering matrix element
$S_{aa}$ is given by solving the single-channel Schr\"{o}dinger
equation for a spherical optical potential. Because the $S$-matrix is
diagonal, the transmission coefficient $T_a$ is defined as a unitarity
deficit
\begin{equation}
  T_{lj} = 1 - |S_{lj,lj}|^2 \ ,
  \label{eq:Tj1}
\end{equation}
where we denote the channel quantum numbers explicitly by the orbital
angular momentum $l$, and spin $j$. We also add a discrete level index
$n$ as $T_{lj}^{(n)}$.  $T_{lj}^{(0)}$ stands for a probability to
form a CN state from the ground state.  The detailed-balance equation
in Eq.~(\ref{eq:HF}) is schematically shown in
Fig.~\ref{fig:transmissionSOM} for the two level case.  $T_{lj}^{(1)}$
stands for a probability to form the same CN from the first excited
state. However, since optical potentials for excited nuclei are
usually unknown, they are replaced by $T_{lj}^{(0)}$ and shift the
energy by the excitation energy $E_x^{(n)}$ to take account of the
energy difference,
\begin{equation}
  T_{lj}^{(n)}(E) \simeq T_{lj}^{(0)}(E - E_x^{(n)}) \ .
  \label{eq:Tj2}
\end{equation}
There exist many phenomenological global optical potentials that are
energy-dependent, and the assumption of Eq.~(\ref{eq:Tj2}) enables all
of the HF codes to perform cross section calculations in wide
energy and target-mass ranges. In fact, almost all of the HF codes generate
$T_{lj}^{(0)}$ on a fixed energy grid before performing a CN
calculation, and interpolate $T_{lj}^{(0)}$ to obtain a required value
at each energy point.

\subsection{Coupled-channels transmission coefficient}

\begin{figure}
  \begin{center}
    \resizebox{0.7\columnwidth}{!}{\includegraphics{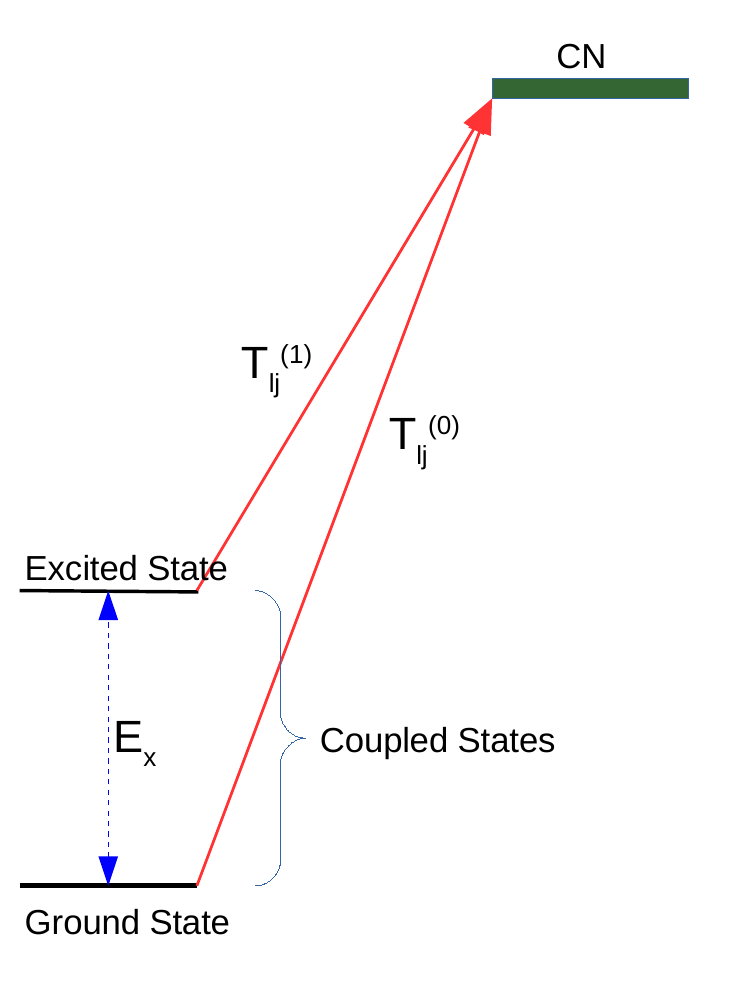}}
   \end{center}
  \caption{In the coupled-channels method, the transmission coefficients 
    $T_{lj}^{(0)}$ and $T_{lj}^{(1)}$ can be derived simultaneously from
    the $S$-matrix.}
  \label{fig:transmissionCC}
\end{figure}

When strongly coupled collective levels are involved in the target
system, Eq.~(\ref{eq:HF}) must be calculated by the transmission
coefficients in the CC formalism, and the $S$-matrix is no longer
diagonal. ECIS and CoH$_3$ calculate generalized transmission
coefficients for all the included states from the CC
$S$-matrix~\cite{Kawano2009}. The time-reversal symmetry of $S$-matrix
yields the transmission coefficient for all of the $n$-th excited state 
simultaneously as
\begin{eqnarray}
  T_{lj}^{(n)} &=&
    \sum_{J\Pi} \sum_a
    {{2s+1}\over{2j_a+1}} g_J \left( 1- \sum_b |S_{ab}^{J\Pi}|^2
    \right) \nonumber \\
  &\times&  \delta_{n_a,n} \delta_{l_a,l} \delta_{j_a,j} \ ,
  \label{eq:Tj3}
\end{eqnarray}
where $J\Pi$ is the total spin and parity. This is shown in
Fig.~\ref{fig:transmissionCC}. The spin-factor $g_J$ is given by
\begin{equation}
  g_J = {{2J+1}\over{(2s+1)(2I_n+1)}} ,
  \label{eq:spinfactor}
\end{equation}
where $s$ is the intrinsic spin of the projectile (= 1/2 for neutron), 
and $I_n$ is the target spin of $n$-th level.
The summation runs over the parity conserved channels,
albeit we omit a trivial parity conservation.
In this expression, cross sections to the directly
coupled channels are eliminated to ensure the sum of $T_{ij}^{(n)}$
gives a correct CN formation cross section from the $n$-th level
\begin{eqnarray}
 \sigma_R^{(n)}
  &=& {{\pi}\over{k_n^2}}
     \sum_{J\Pi} \sum_a \delta_{n_a,n} g_J
     \left(1-\sum_b |S_{ab}^{J\Pi}|^2\right) \nonumber \\
  &=& {{\pi}\over{k_n^2}} \sum_{lj} {{2j+1}\over{2s+1}} T_{lj}^{(n)}  .
 \label{eq:sigreacT}
\end{eqnarray}
Now the off-diagonal elements in the $S$-matrix are effectively
eliminated. By substituting $T_{lj}^{(n)}$ into Eq.~(\ref{eq:HF}), the HF
cross section is determined in terms of the detailed balance. This
formulation is, however, valid only when the width fluctuation
correction is not so significant, because the off-diagonal elements in $S$ to
calculate $W_{ab}$ are ignored. We will discuss this later.

The HF codes, exept for ECIS and CoH$_3$ to our knowledge, simplify
Eq.~(\ref{eq:Tj3}) by applying the approximation of
Eq.~(\ref{eq:Tj2}), namely $T_{lj}^{(n)}(E) \simeq
T_{lj}^{(0)}(E-E_x^{(n)})$, hence all the CC $S$-matrix properties are
lost. To examine this approximation, we calculate transmission
coefficients for the excited states $T_{lj}^{(n)}(E)$ and compare with
those for the energy-shifted ground state
$T_{lj}^{(0)}(E-E_x^{(n)})$. The comparison includes two cases of
rotational band head; the target ground state spin is zero, and it is
half-integer. The first example is for the fast-neutron induced
reaction on $^{238}$U, in which we couple 5 levels (0, 45, 148, 307,
and 518~keV, from $0^+$ to $8^+$) in the ground state rotational
band. The optical potential of Soukhovitskii et
al.~\cite{Soukhovitskii2004} is used.  A natural choice for the second
case, the half-integer ground state spin, is $^{239}$Pu. However, its
large fission cross sections at low energies blur the difference
coming from the definition of $T_{lj}$. Instead, we adopt $^{169}$Tm
that has a similar level structure to
$^{239}$Pu~\cite{Kawano2009}. The coupled levels are 0, 8.4, 118, 139,
and 332~keV from $(1/2)^+$ to $(9/2)^+$, and the optical potential of
Kunieda et al.~\cite{Kunieda2007} is employed.

\begin{figure}
  \begin{center}
    \resizebox{0.85\columnwidth}{!}{\includegraphics{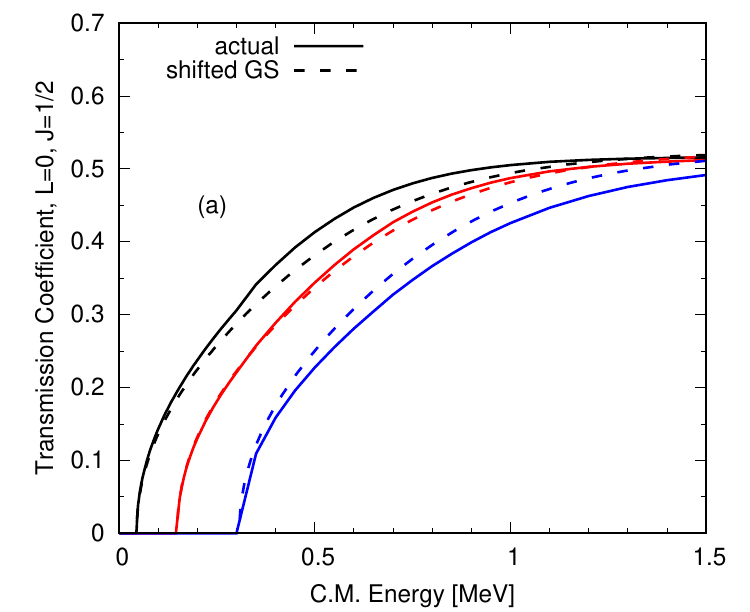}} \\
    \resizebox{0.85\columnwidth}{!}{\includegraphics{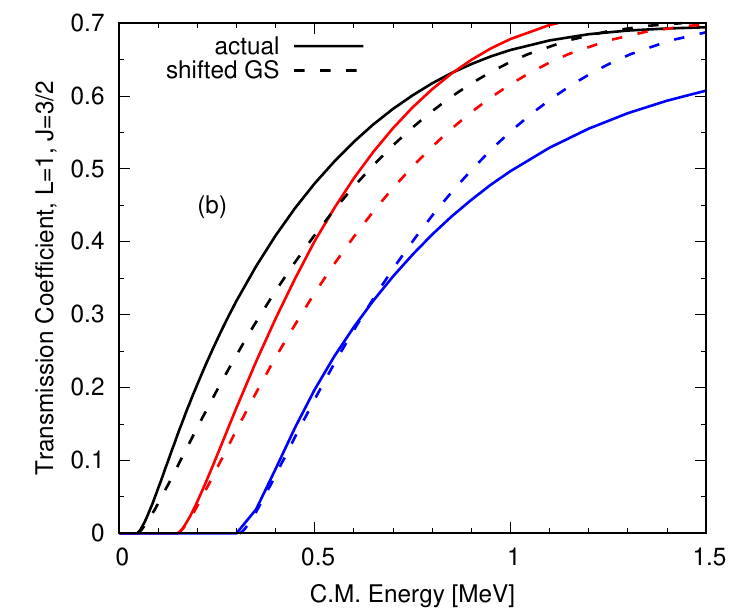}} \\
    \resizebox{0.85\columnwidth}{!}{\includegraphics{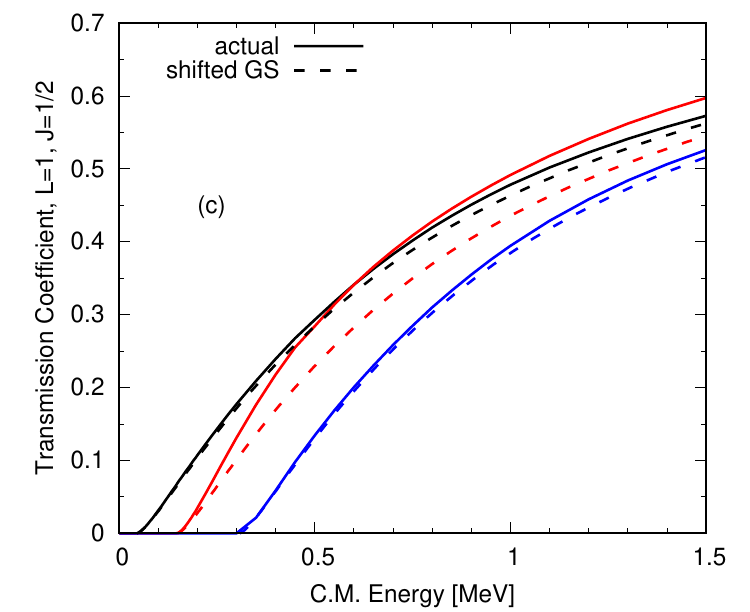}}
 \end{center}
 \caption{Calculated transmission coefficients for the first and
    second coupled levels of $^{238}$U; (a) for $(l,j)=(0,1/2)$, (b)
    $(1,3/2)$, and (c) $(1,1/2)$. The solid curves are the actual
    transmission coefficient given by the CC $S$-matrix. The dashed
    curves are approximation by the ground state transmission
    coefficient shifted by the level excitation energies. The black,
    red, and blue curves are for the first, second, and third levels.
    The corresponding level should also be distinguished
    by the shifted threshold energies.}
 \label{fig:u238trans}
\end{figure}

\begin{figure}
  \begin{center}
    \resizebox{0.85\columnwidth}{!}{\includegraphics{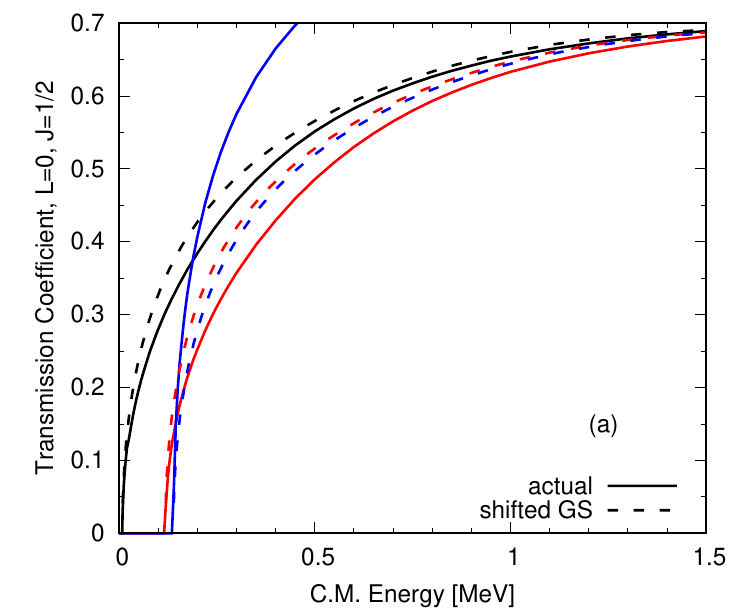}} \\
    \resizebox{0.85\columnwidth}{!}{\includegraphics{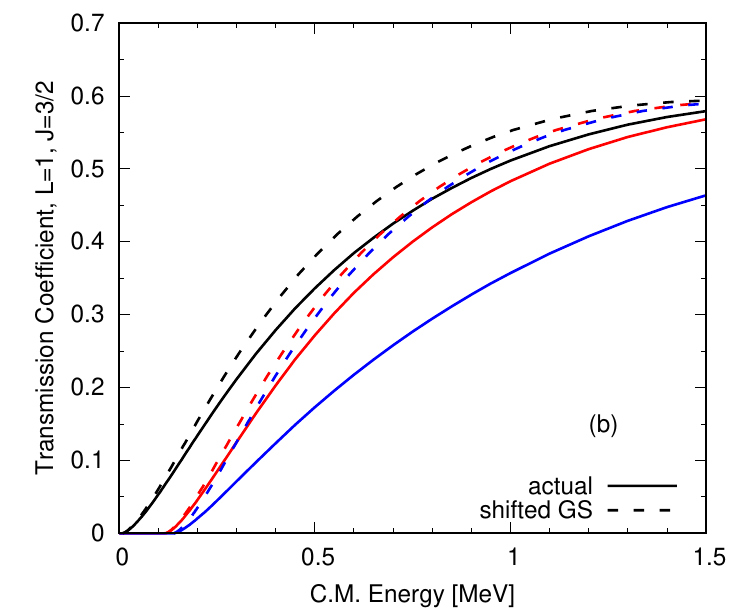}} \\
    \resizebox{0.85\columnwidth}{!}{\includegraphics{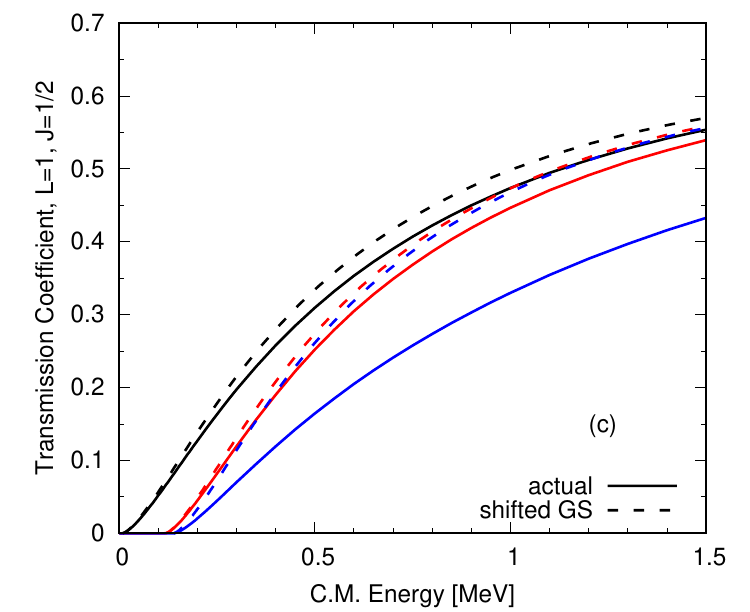}}
 \end{center}
  \caption{Same as Fig.~\ref{fig:u238trans} but for $^{169}$Tm.}
  \label{fig:tm169trans}
\end{figure}

Figure~\ref{fig:u238trans} shows the difference in the $l=0$ and 1
transmission coefficients for $^{238}$U.  The approximation by
$T_{lj}^{(0)}$ seems to be reasonable for the $s$-wave transmission
coefficient, while the difference reaches about 10\% for the $p$-wave
case. The calculations for $^{169}$Tm, shown in
Fig.~\ref{fig:tm169trans}, show the opposite tendency; a notable
difference appears in the $s$-wave. It is difficult to draw a general
conclusion by these limited examples. However, it is obvious that the
calculated cross sections by feeding these transmission coefficients
into the statistical HF theory are no longer equivalent, and the
energy-shifted transmission coefficient inflates uncertainty in the
calculated results.

\subsection{Generalized transmission coefficient in compound inelastic scattering calculation}

To see the actual impact of the generalized transmission coefficients
on the cross section calculation, we have to calculate the HF equation
with these actual/approximated transmission
coefficients. Unfortunately, this is not so easy in general, because
it requires extensive modification to the computer programs.  Instead,
we made an {\it ad-hoc} modification to CoH$_3$ to test this.  For a
100-keV neutron induced reaction on $^{238}$U, CoH$_3$ with the
generalized transmission coefficient gives the inelastic scattering
cross section of 553~mb to the 45~keV $2^+$ level.  At this energy, an
equivalent center-of-mass (CMS) energy to the first 45~keV level is
$100 A/(A+1) - 44.9 = 54.7$~keV. We calculate $T_{lj}^{(0)}$ at
$E_{\rm CMS}=54.7$~keV, and replace $T_{lj}^{(1)}$ by these values.
The calculated inelastic scattering cross section is 488~mb, which is
9\% smaller than the generalized transmission case, and closer to the
evaluated value in JENDL-4~\cite{Iwamoto2007,JENDL4} of 461~mb.  This
observation is consistent with the larger $p$-wave transmission
coefficient as shown in Fig.~\ref{fig:u238trans} (b).

In the past, a code comparison was carried out~\cite{Capote2017} by
including EMPIRE~\cite{INDC0603}, TALYS~\cite{Koning2012,Koning2008},
CCONE~\cite{Iwamoto2016}, and CoH$_3$.  The result revealed that the
inelastic scattering cross section by CoH$_3$ tends to be higher than
those by the other codes for the $^{238}$U case at low energies.  The
difference is about 10\% at 100~keV, and this confirms our numerical
exercise here; the approximation by the ground state transmission
coefficient systematically underestimates the inelastic scattering
cross section for the $^{238}$U case.

\subsection{Transmission coefficient for uncoupled channels}

\begin{figure}
  \begin{center}
    \resizebox{0.7\columnwidth}{!}{\includegraphics{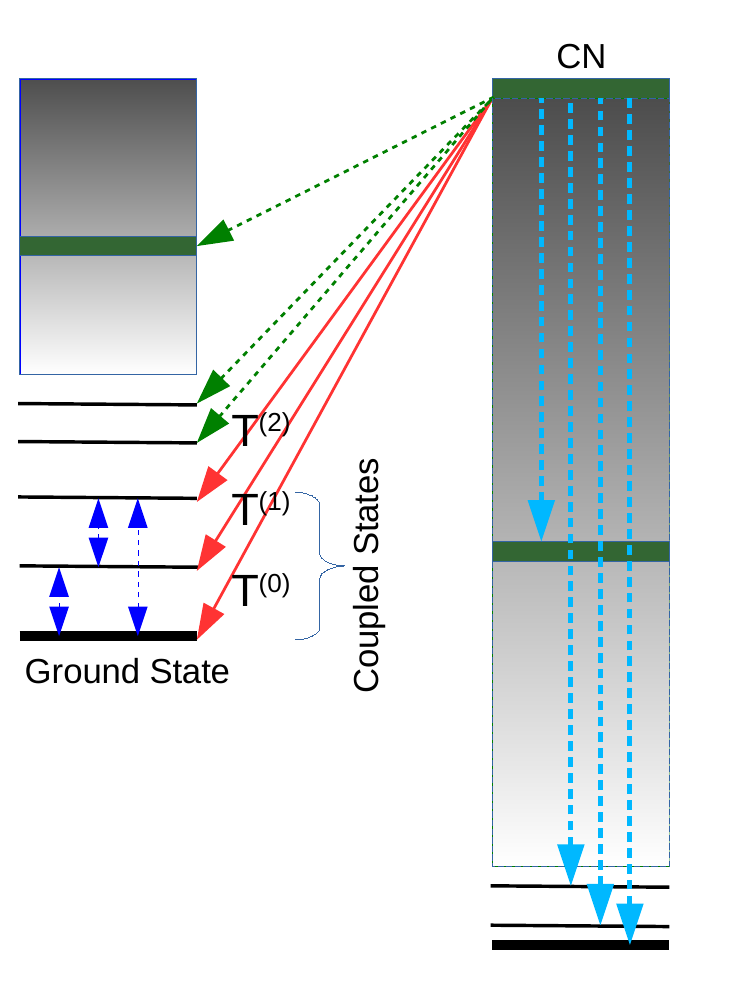}}
   \end{center}
  \caption{All the possible decay channels from a compound state are schematically shown by 
    the arrows. The solid arrows are the transmission coefficients by the CC model.
    All the other channels shown by the dotted arrows are uncoupled levels.}
  \label{fig:transmissionAll}
\end{figure}

There might be many uncoupled levels involved in actual CN
calculations, as schematically shown in
Fig.~\ref{fig:transmissionAll}. In the target nucleus, there are
uncoupled discrete levels up to some critical energy, then a level
density model is used to discretize the continuum above there. The
transmission coefficients to these states are calculated by the
single-channel case.

The formed CN state can decay by emitting a charged-particle,
$\gamma$-ray or fission. The charged-particle transmission
coefficients are basically the same as the uncoupled neutron channel
case, except for the Coulomb interaction.  The transmission
coefficients for the $\gamma$ decay are calculated by applying the
giant dipole resonance (GDR) model~\cite{Goriely2019}, where GDR
parameters derived from experimental data or theoretically predicted
are often tabulated~\cite{Goriely2019,Kawano2020}. Although there are
a large number of final states available after a $\gamma$-ray
emission, these probabilities are very small compare to the neutron
transmission coefficients. Often it is good enough to lump the
$\gamma$-ray channels into a single $\gamma$-ray transmission
coefficient as
\begin{equation}
  T_\gamma = \sum_{XL}\int_0^{E_n+S_n} T_{XL}(E_\gamma) \rho(E_x) dE_x \ ,
  \label{eq:Tgamma}
\end{equation}
where $S_n$ is the neutron separation energy, $X$ stands for the type
of radiation (E: electric, M: magnetic), $L$ is the multipolarity, and
$\rho(E_x)$ is the level density at excitation energy $E_x = E_n+
S_n-E_\gamma$. When the final state is in a discrete level, the
integration in Eq.~(\ref{eq:Tgamma}) is replaced by an appropriate
summation.

When the CN fissions, the simplest expression of the fission
transmission coefficients is the WKB approximation to the inverted
parabola shape of fission barriers proposed by Hill and
Wheeler~\cite{Hill1953}. Albeit it is know that this form has an issue
to reproduce experimental fission cross sections, this is beyond the
scope of current paper, and we do not discuss it further.
The fission takes place though many states on top of the fission barrier,
so that it is convenient to lump these partial fission probabilities
into the fission transmission coefficient $T_f$.

\subsection{Engelbrecht-Weidenm\"{u}ller transformation for width fluctuation correction}

The width fluctuation correction factor $W_{ab}$ consists of the
elastic enhancement factor and the actual width fluctuation correction
factor~\cite{Moldauer1975a}. A convenient definition is suggested by
Hilaire, Lagrange, and Koning~\cite{Hilaire2003}, which is to define
as a ratio to the pure HF cross section as in Eq.~(\ref{eq:HF}). When
we ignore the channel coupling effect, $W_{ab}$ in
Eq.~(\ref{eq:MoldauerW}) can be calculated by the generalized
transmission coefficients in Eq.~(\ref{eq:Tj3}). However, we cannot
employ this prescription when a strong channel-coupling results in
non-negligible off-diagonal elements in $S$. Instead, we perform the
Engelbrecht-Weidenm\"uller (EW) transformation~\cite{Engelbrecht1973}
to correctly eliminate the off-diagonal elements.

Satchler's transmission matrix~\cite{Satchler1963} is defined by
the CC $S$-matrix as
\begin{equation}
  P_{ab} = \delta_{ab} - \sum_c S_{ac} S_{bc}^*  \ .
  \label{eq:Pmatrix}
\end{equation}
Since $P$ is Hermitian, we can diagonalize this by a unitary
transformation~\cite{Engelbrecht1973}
\begin{equation}
  (UPU^\dag)_{\alpha\beta} = \delta_{\alpha\beta} p_\alpha,
  \qquad 0 \le p_\alpha \le 1 \ ,
\end{equation}
where $\alpha$ and $\beta$ are the channel indices in the diagonalized
space. The diagonal element $p_\alpha$ is the new transmission coefficient,
because the $S$-matrix is also diagonalized as
\begin{equation}
  \tilde{S} = U S U^T \ ,
\end{equation}
which defines the single-channel transmission coefficient
\begin{equation}
  T_\alpha = 1 - \left|\tilde{S}_{\alpha\alpha}\right|^2  = p_\alpha \ .
  \label{eq:eigenvalue}
\end{equation}

We now calculate the width fluctuation in the diagonal channel space, then
transform back to the cross-section space by~\cite{Hofmann1975}
\begin{eqnarray}
  \sigma_{ab}
  &=& \sum_\alpha |U_{\alpha a}|^2 |U_{\alpha b}|^2 \sigma_{\alpha\alpha} \nonumber\\
  &+& \sum_{\alpha\neq\beta} U_{\alpha a}^* U_{\beta b}^*
       \left(
         U_{\alpha a} U_{\beta b} + U_{\beta a} U_{\alpha b}
       \right) \sigma_{\alpha\beta} \nonumber\\
  &+& \sum_{\alpha\neq\beta} U_{\alpha a}^* U_{\alpha b}^* U_{\beta a} U_{\beta b}
         \ave{\tilde{S}_{\alpha\alpha} \tilde{S}_{\beta\beta}^*} .
  \label{eq:backtrans2}
\end{eqnarray}
$\sigma_{\alpha\alpha}$ and $\sigma_{\alpha\beta}$ are the width
fluctuation corrected cross section
with the transmission coefficient of $p_\alpha$.
The last term $\ave{\tilde{S}_{\alpha\alpha} \tilde{S}_{\beta\beta}^*}$ was
evaluated by applying the Monte Carlo technique to GOE~\cite{Kawano2016},
\begin{equation}
  \overline{\tilde{S}_{\alpha\alpha} \tilde{S}_{\beta\beta}^*}
  \simeq e^{i(\phi_\alpha - \phi_\beta)}
    \left( \frac{2}{\nu_\alpha} - 1 \right)^{1/2}
    \left( \frac{2}{\nu_\beta} - 1 \right)^{1/2}
    \sigma_{\alpha\beta}  ,\\
  \label{eq:Sab2}
\end{equation}
where $\phi_\alpha =\tan^{-1} \tilde{S}_{\alpha\alpha}$. Here we
replaced the energy average $\ave{*}$ by the ensemble average
$\overline{*}$.  Applying the GOE model to the channel
degree-of-freedom $\nu_\alpha$~\cite{Kawano2015}, the HF cross section
with the width fluctuation correction is fully characterized in the CC
framework.

When uncoupeld-channels, such as the inelastic scattering to the higher levels,
$\gamma$-decay and fission channels, exist, the transmission matrix has
these sub-space
\begin{equation}
   P = \left(
 \begin{array}{cccc}
   P_{\rm C} &     &         & \\
             & T_n &         & \\
             &     & T_\gamma & \\
             &     &         & T_f \\
 \end{array} \right),
\end{equation}
where $P_{\rm C}$ is the coupled channels $P$ matrix in
Eq.~(\ref{eq:Pmatrix}). Because $T_n$, $T_\gamma$, and $T_f$ are still
diagonal, the unitary transformation is only applied to $P_{\rm C}$.
The uncoupled cross section is calculated by~\cite{Kawano2016}
\begin{equation}
  \sigma_{ab} = \sum_{\alpha} |U_{\alpha a}|^2 \sigma_{\alpha\beta} \delta_{\beta b} \ .
  \label{eq:uncoupled}
\end{equation}

Here we take $^{238}$U and $^{169}$Tm as examples again. The width
fluctuation correction is defined as a ratio to the HF cross section,
and we calculate two cases; (a) the width fluctuation factor $W_{ab}$
in Eq.~(\ref{eq:MoldauerW}) is calculated by using the generalized
transmission coefficient of Eq.~(\ref{eq:Tj3}), and the off-diagonal
elements in the $S$-matrix are ignored, and (b) the full EW
transformation is performed.

\begin{figure}
  \begin{center}
    \resizebox{0.85\columnwidth}{!}{\includegraphics{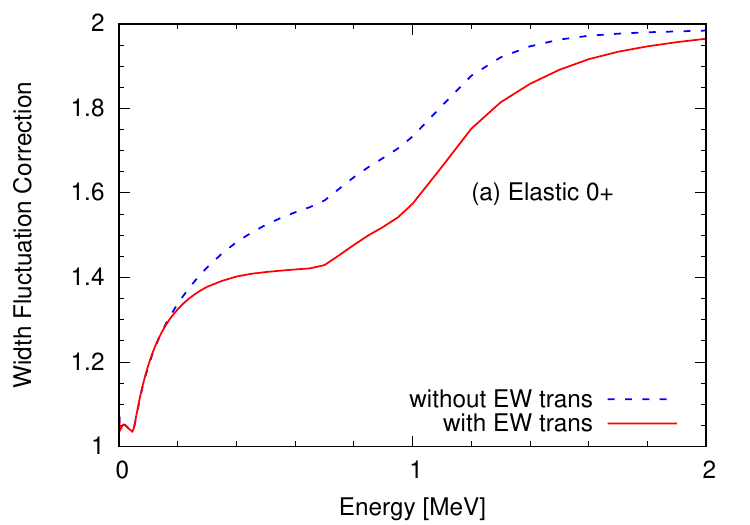}} \\
    \resizebox{0.85\columnwidth}{!}{\includegraphics{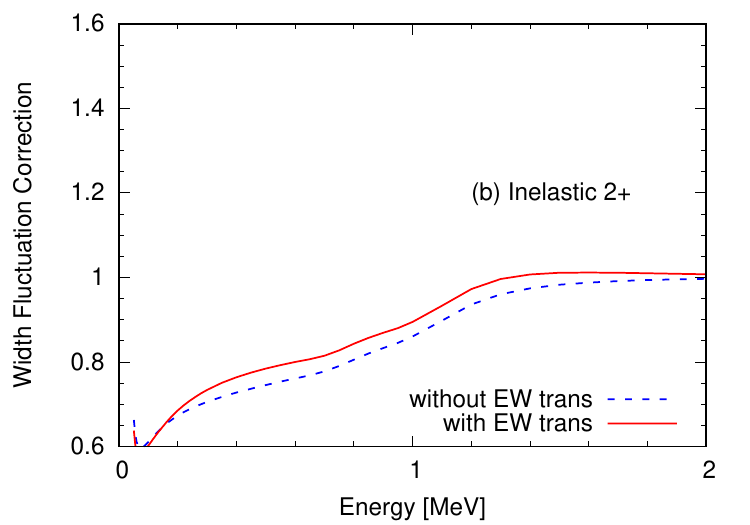}} \\
    \resizebox{0.85\columnwidth}{!}{\includegraphics{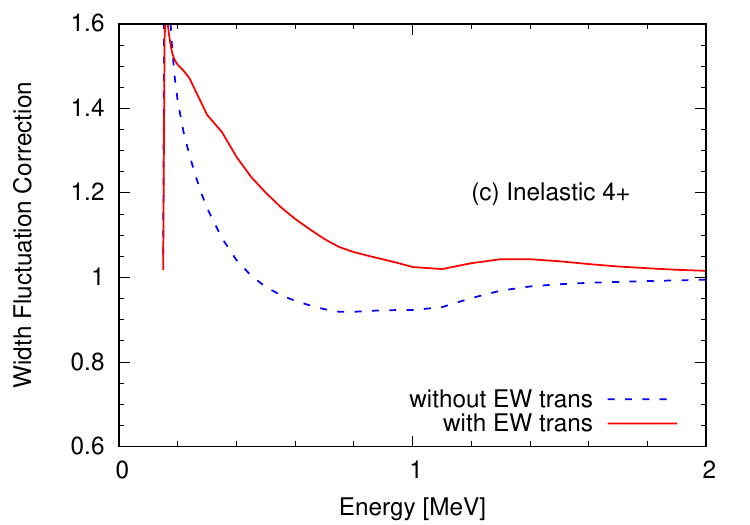}}
 \end{center}
 \caption{Width fluctuation corrections for $^{238}$U
    defined as a ratio to the Hauser-Feshbach cross section.
    The panels (a), (b), and (c) are the compound elastic, inelastic to 
    the first, and second levels for the neutron-induced reaction on $^{238}$U.
    The solid curves are calculated by performing the Engelbrecht-Weidenm\"{u}ller (EW)
    transformation, and the dashed curves are without the EW transformation.}
  \label{fig:u238wf}
\end{figure}

It is known that an asymptotic value of $W_{aa}$ (elastic enhancement
factor) is 2 when all the channels are equivalent.
Figure~\ref{fig:u238wf} (a), which is the calculated $W_{aa}$
for $^{238}$U, shows this behavior, but the EW
transformation slightly deters $W_{aa}$ from approaching the
asymptote. The weaker elastic enhancement results in increase in the
inelastic scattering channels. This is also demonstrated by the Monte
Carlo simulation for the GOE scattering matrix when direct reaction
components are involved~\cite{Kawano2015}. In other words, the
directly coupled channels squeeze the elastic scattering channel due
to constraint by the $S$-matrix unitarity, hence the enhancement in
the elastic channel will have less influence on the other channels.

\begin{figure}
  \begin{center}
    \resizebox{0.85\columnwidth}{!}{\includegraphics{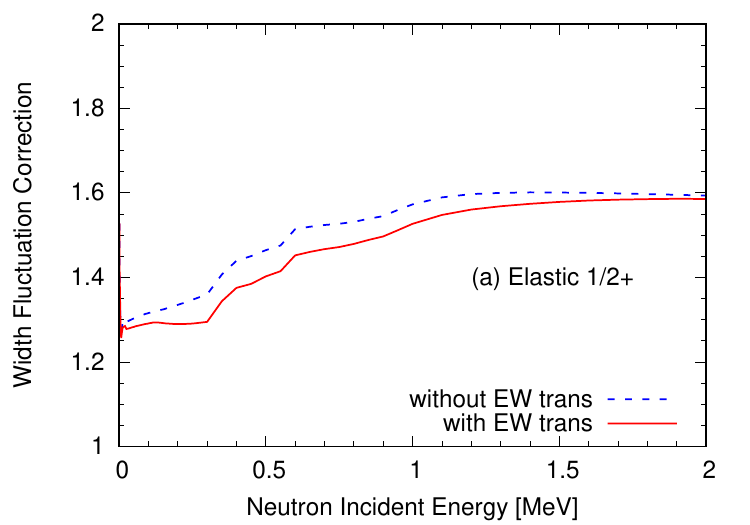}} \\
    \resizebox{0.85\columnwidth}{!}{\includegraphics{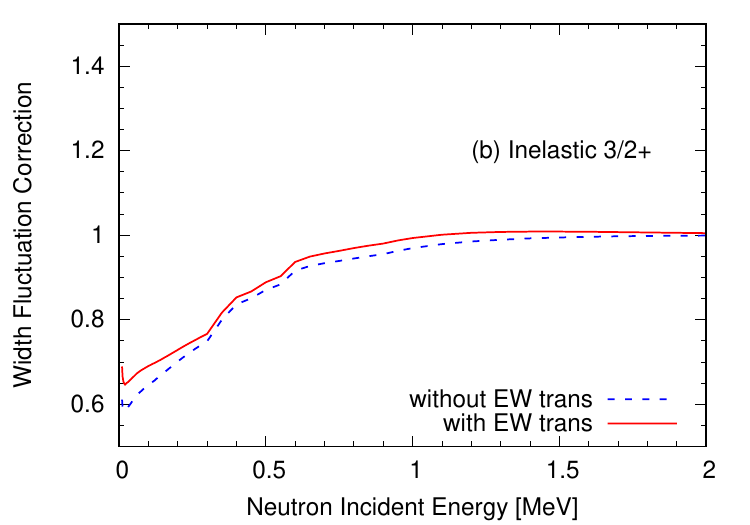}} \\
    \resizebox{0.85\columnwidth}{!}{\includegraphics{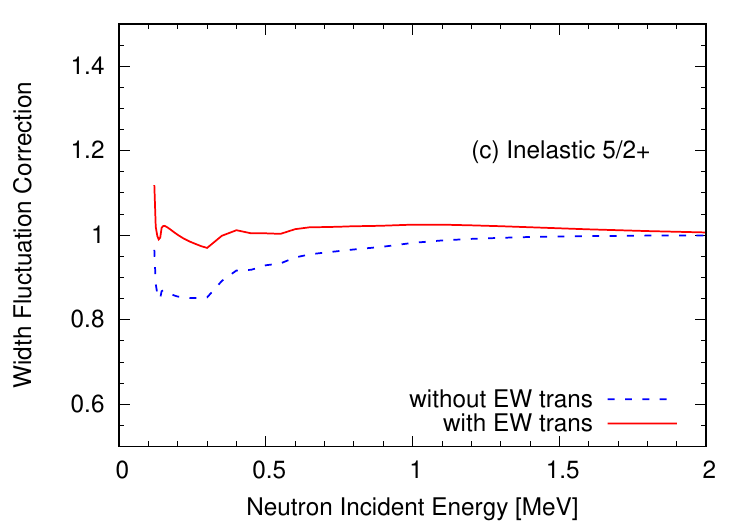}}
 \end{center}
  \caption{Same as Fig.~\ref{fig:u238wf} but for $^{169}$Tm.}
  \label{fig:tm169wf}
\end{figure}

In the case of $^{169}$Tm, shown in Fig.~\ref{fig:tm169wf}, the
asymptotic value of $W_{aa}$ does not reach 2 but stays about
1.6. This might be because the $s$-wave transmission coefficient for
the second excited state is very different from the other
channels. Because the number of channels is larger than the $^{238}$U
case, the EW transformation less impacts the CN calculations. In
addition, other uncoupled channels, e.g. radiative capture and
fission channels if exist, further mitigate the elastic enhancement
effect. Therefore the EW transformation is mostly important for
rotating even-even nuclei with large deformation. Having said that,
the difference seen in Fig.~\ref{fig:tm169wf} implies an inherent
deficiency in the simplified HF calculations widely adopted nowadays.

\section{Application}
\label{sec:application}
In reality, the EW transformation does not modifies the calculated
cross sections so largely. As seen in Fig.~\ref{fig:u238wf}, the
difference between the EW and single-channel $W_{ab}$ cases is at
most 15\%. Such a difference may occur due to other uncertain inputs to
the calculation. One of the most crucial model parameters is the
optical potential. The optical potential parameters are often obtained
phenomenologically by fitting to experimental elastic scattering and
total cross sections. In general, similar quality of data fitting can be
achieved by different potential parameters, while they may have
slightly different partial wave contributions. Fluctuation in the
partial wave contribution is sometimes visible in the inelastic
scattering cross sections, where limited numbers of partial waves are
involved.

\begin{figure}
  \begin{center}
    \resizebox{\columnwidth}{!}{\includegraphics{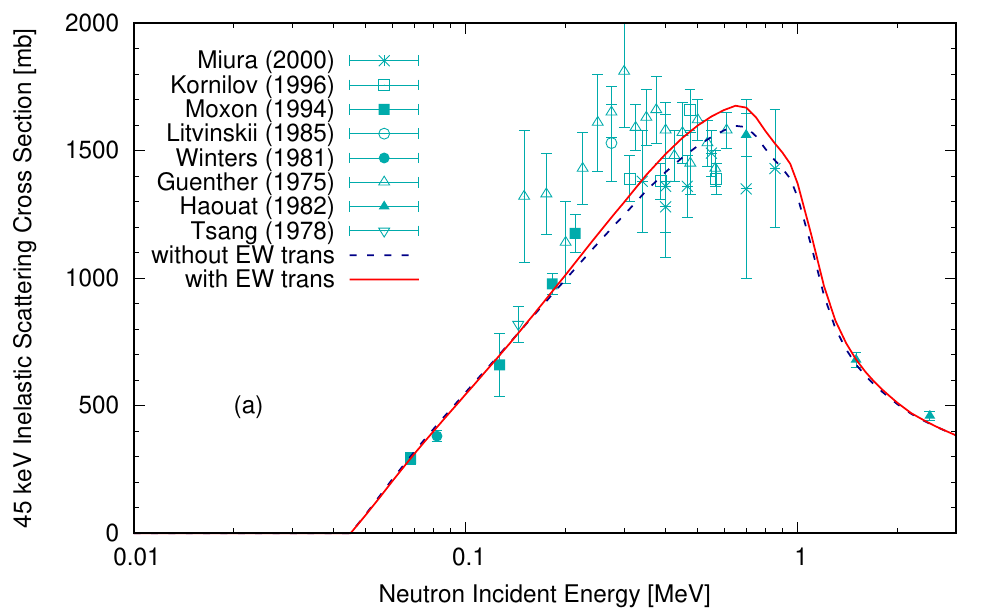}}\\
    \resizebox{\columnwidth}{!}{\includegraphics{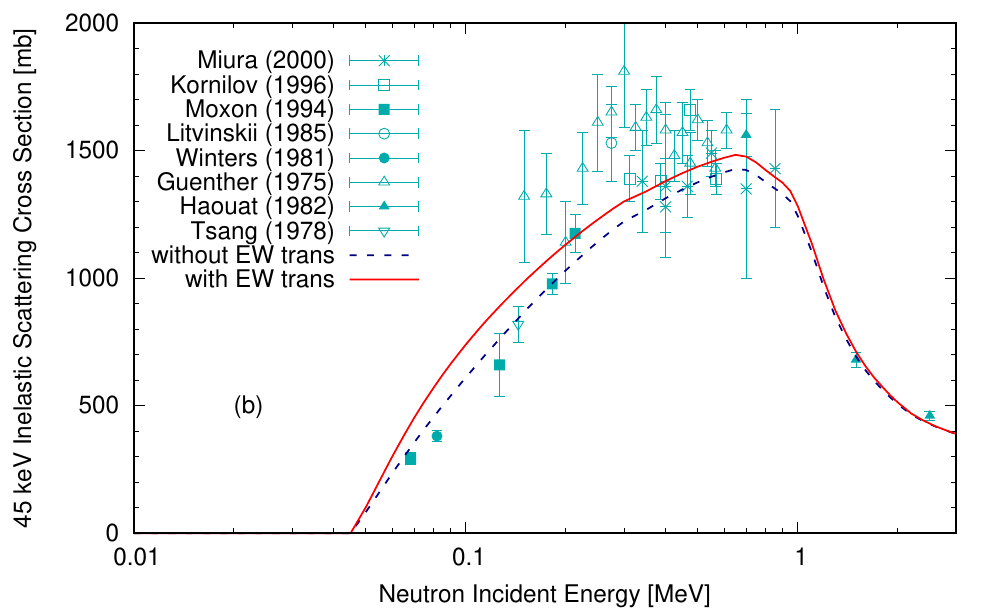}}
 \end{center}
  \caption{Comparison of calculated inelastic scattering cross section to the first 
           45~keV level of $^{238}$U. The top panel (a) is the case of 
           2004 Soukhovitskii potential~\cite{Soukhovitskii2004}, 
           and (b) is the 2005 Soukhovitskii potential~\cite{Soukhovitskii2005}}
  \label{fig:u238inel}
\end{figure}

Figure~\ref{fig:u238inel} (a) shows a comparison of the calculated
inelastic scattering cross section to the first 45~keV level of
$^{238}$U with available experimental data of Miura et
al.~\cite{Miura2000}, Kornilov and Kagalenko~\cite{Kornilov1995},
Moxon et al.~\cite{Moxon1994}, Litvinskii et
al.~\cite{Litvinskii1990}, Winters et al.~\cite{Winters1981}, Guenther
et al.~\cite{ANLNDM16}, Haouat et al.~\cite{Haouat1982}, and Tsang and
Brugger~\cite{Tsang1978}. We performed this calculation with the CC
optical model potential of Soukhovitskii et
al.~\cite{Soukhovitskii2004}. While the measurements are largely
scattered in the hundreds keV region, the EW transformation moves the
calculation into a preferable direction. However, the enhancement due
to the EW transformation is rather modest, which is also seen in
Fig.~\ref{fig:u238wf} (b).

When we switch the optical potential into the updated Soukhovitskii
potential~\cite{Soukhovitskii2005}, the EW transformation becomes
noticeable as shown in Fig.~\ref{fig:u238inel} (b). In this case the
enhancement is visible in the wider energy range. Of course it is not
rational to verify an optical potential by applying it to the
statistical model, as ambiguity caused by other model inputs 
persists. In the CC formalism, calculated cross sections are also
influenced by the coupling scheme~\cite{Dietrich2012}.  Despite other
available optical potentials for $^{238}$U may provide different
excitation functions of 45-keV level, we may say generally that the EW
transformation increases the 45-keV level cross section due to
the hindered elastic enhancement.

\begin{figure}
  \begin{center}
    \resizebox{\columnwidth}{!}{\includegraphics{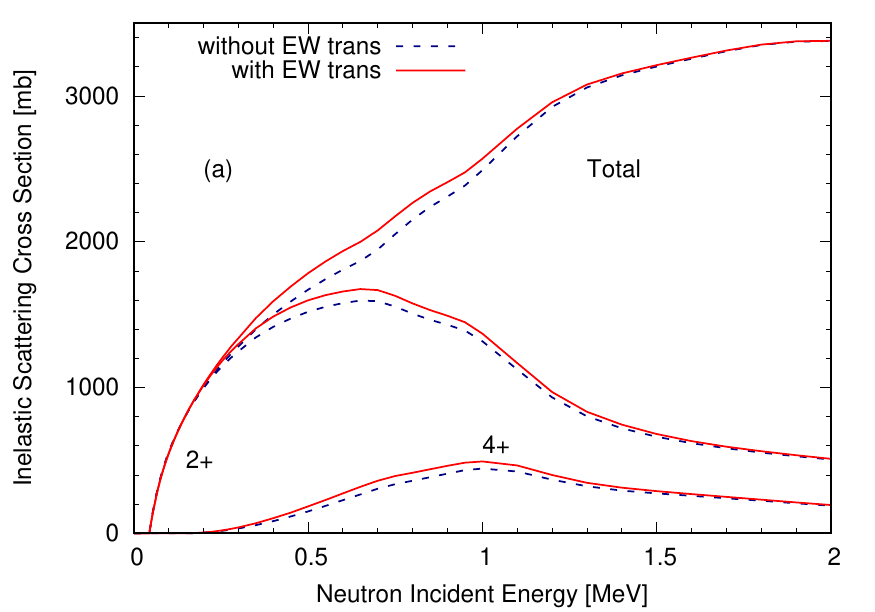}}\\
    \resizebox{\columnwidth}{!}{\includegraphics{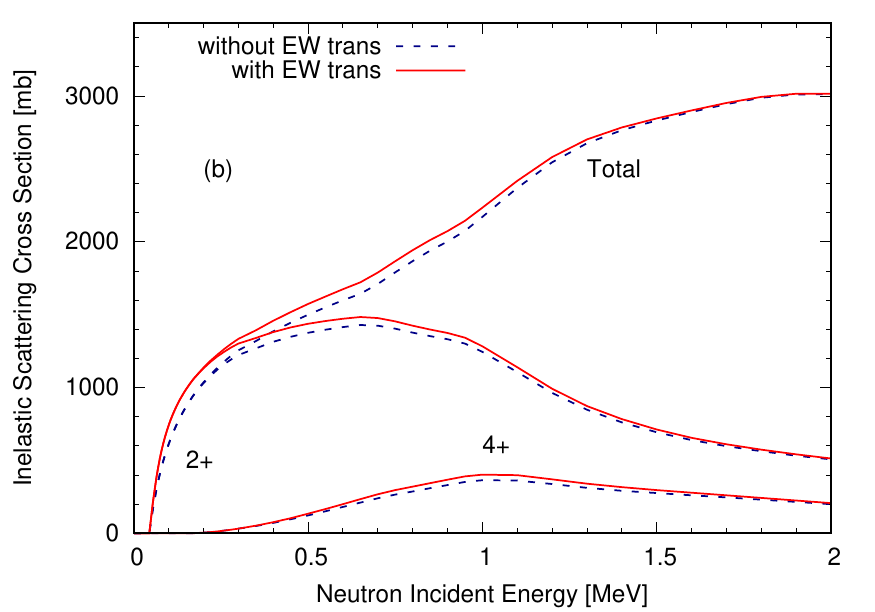}}
 \end{center}
  \caption{The inelastic scattering cross section to the 45~keV and 
           148~keV levels, and the total inelastic scattering cross section
           of $^{238}$U. The top panel (a) is for the 2004
           Soukhovitskii potential~\cite{Soukhovitskii2004}, 
           and (b) is the 2005 Soukhovitskii potential~\cite{Soukhovitskii2005}}
  \label{fig:u238ineltot}
\end{figure}

The increase in the inelastic scattering of the 148-keV level, as well
as the total inelastic scattering cross section, is shown in
Fig.~\ref{fig:u238ineltot}. The $2^+$ cross sections are the same as
those in Fig.~\ref{fig:u238inel}. Since the relative magnitude of the
$4^+$ level cross section is smaller than the $2^+$ level, this has a
minor impact on the total inelastic scattering cross section. This is
also true for the higher spin states ($6^+, 8^+ \ldots$)

It might be worth reminding that these ``without EW'' cases employ the
generalized transmission coefficient to calculate both the HF cross
section and the $W_{ab}$ factor. When one adopts a conventional
prescription of $T^{(n)}(E) \simeq T^{(0)}(E - E_x^{(n)})$, the
calculated inelastic scattering cross section would be further lower
than the no-EW case.  Evidently this approximation cannot be justified
anymore when the nuclear deformation plays an important role. Use of
the generalized transmission calculation is still approximated, albeit
it mitigates this deficiency to some extent, and afford us not so
heavy computation. However, as we demonstrated the quantitative
deficiencies in the approximations and simplifications made so far, we
should consider implementing the EW transformation in the HF model
codes for better prediction of nuclear reaction cross sections for
deformed nuclei.

A remaining complication is the angular distributions of scattered
particles in the CN process. The differential cross section is
expanded by the Legendre polynomials in the Blatt-Biedenharn
formalism~\cite{Blatt1952},
\begin{equation}
  \left(\frac{d\sigma}{d\Omega}\right)_{ab}
   = \sum_L B_L P_L(\cos \theta_b) \ ,
 \label{eq:CNleg}
\end{equation}
where the scattering angle $\theta_b$ is in the center-of-mass system.
A full expression of the $B_L$ coefficients is given in the
single-channel width fluctuation case~\cite{Moldauer1964,Kim2020}.
However, a complete formulation of the $B_L$ coefficient becomes very
difficult to calculate when the EW transformation is performed.
Alternatively, we can apply the generalized transmission coefficients
without the EW transformation for calculating $B_L$. This is roughly
the Legendre coefficients in the HF case $B_L^{\rm HF}$ times the
width fluctuation correction factor $W_{ab}$, but more correction
terms are involved~\cite{Kim2020}.

\section{Conclusion}
\label{sec:conclusion}
We presented a general formulation of the statistical Hauser-Feshbach
(HF) theory with width fluctuation correction for a deformed nucleus,
and applied to the low-energy neutron induced reactions on $^{238}$U
and $^{169}$Tm. The main difference between the conventional HF model
is; (a) we calculate generalized transmission coefficients from the
coupled-channels (CC) $S$-matrix, and (b) the width fluctuation
calculation is performed in the diagonalized channel space, which is
the so-called Engelbrecht-Weidenm\"{u}ller (EW)
transformation. Whereas these ingredients were already implemented
into J.~Raynal's coupled-channels code ECIS, the coupled-channels HF code,
CoH$_3$, offers more general functionality for calculating nuclear
reactions at low energies. We demonstrated that both the generalized
transmission coefficients and the EW transformation increase the
neutron inelastic scattering cross section when strongly coupled
direct reaction channels exist. This happens due to the fact that
contributions from each partial wave are different, and that
constraints by the unitarity of $S$-matrix is somewhat relaxed. The HF
nuclear reaction calculation codes currently available in the market
often simplify the deformed nucleus calculations by assuming a nuclear
deformation effect is negligible. Our numerical calculations for a few
examples evidently demonstrated that such the simplification results
in underestimation of the inelastic scattering cross sections.

\begin{acknowledgement}
The author is grateful to E. Bauge, S. Hilaire, and P. Chau of CEA
Bruy\`{e}res-le-Ch\^{a}tel and P. Talou of LANL for encouraging this
work.  This work was carried out under the auspices of the National
Nuclear Security Administration of the U.S. Department of Energy at
Los Alamos National Laboratory under Contract No. 89233218CNA000001.
\end{acknowledgement}


\end{document}